\documentclass[12pt]{iopart}
\usepackage{iopams}
\usepackage{cite}
\usepackage{cleveref}
\usepackage{graphicx}
\usepackage{array,multirow}
\usepackage{dcolumn}% Align table columns on decimal point
\begin{document}

\title{Entanglement and correlation in two-nucleon systems}

\author{A T Kruppa$^1$, J Kov\'acs$^1$, P Salamon$^{1}$, {\" O} Legeza$^{2}$ }

\address{$^1$Hungarian Academy of Sciences Institute for Nuclear Physics, Debrecen, Hungary}
\address{$^2$Wigner Research Centre for Physics, H-1525 Budapest, Hungary}

\begin{abstract}
We examine the mode entanglement and correlation of two fermionic particles.  We study the one- and two-mode entropy and a global characteristic, the one-body entanglement entropy. We consider not only angular momentum coupled states with single configuration but use the configuration interaction method. With the help of the Slater decomposition, we derive analytical expressions for the entanglement measures. We show that when the total angular momentum is zero specific single configurations describe maximally entangled states. 
It turns out that for a finite number of associated modes the  one- and two-mode entropies have identical values. 
In the shell model framework, we numerically study two valence neutrons in the $sd$ shell. The one-body entanglement entropy of the ground state is close to the maximal value and the associated modes have the largest mutual information.
\end{abstract}

\maketitle

\section{Introduction}

Recently, the study of non-classical correlations between individual subsystems has gained importance in several research areas. 
Maybe the most notable form of non-classical correlation is entanglement~\cite{hor09,ami08}. 
The significance of these can be approached from several viewpoints.
On the one hand, the experimental observation of non-classical correlations predicted by quantum theory represents a constraint on the theories to be used. 
On the other hand, entanglement provides a resource for communication and computing that goes beyond the possibilities of classical physics \cite{nie00}. 

The entanglement of subsystems is a highly diverse problem, raising several open questions and approaches.
One of the fundamental issues is the choice of subsystems to be considered in the studies. 
In the case of distinguishable particles,  the notion of the entanglement is based on the structure of the tensor product of Hilbert spaces of the subsystems. 
This type of entanglement is very thoroughly investigated \cite{hor09}.

The studies on the entanglement of identical particles have led to tough conceptual questions that have been studied for years.
In the case of  identical fermions the system is described by the antisymmetric part of the Fock space, however  
the decomposition into particle subsystems  does not correspond
to tensor product structure of the Fock space.
To overcome this problem, the notion of the mode entanglement was introduced \cite{zan02,git02,shi03,leg03}.
In the second quantized formalism,  the mode creation and annihilation operators generate 
the algebra of observables. The subsystems are defined in terms of subalgebras \cite{ban07,ben14,ben14b} and the partial trace operation is replaced by restriction of the quantum state to a subalgebra.
With a carefully defined  partial trace operation  on the Fock space \cite{fri13,fri16} one can define the density operator of the subsystem.
The great advantage of the mode entanglement is that the modes can form a feasible, well-defined subsystems even for indistinguishable particles \cite{fri13,ben14,dal17}. Recently entanglement investigations  \cite{bal13a,bal13,ben16} are carried out in the framework of the algebraic quantum mechanics where the physical observables are described by a $C^*$ algebra. Nevertheless both approaches mode entanglement \cite{sha19,deb17,gig17,gig16,szal20} and 
particle based studies \cite{eck02,lev05,ghi04,pas01,sch01,maj16,kwa17,din20a} are present to describe  quantum correlations in fermionic systems. 

In our work, we apply the mode entanglement characterization of the quantum states.
In atomic physics, the quantitative characterization of the entanglement has already been studied in several models \cite{tic11}. 
The application of quantum information concepts has also proved to be extremely useful in the study of the chemical bond \cite{sza15,ste18,sza17}.
The study of the mode entanglement in nuclei is an almost entirely unexplored area. Although initial steps have already been taken to study entanglement in nuclear physics models. 
The investigations of the non-classical correlations in the Lipkin model \cite{tul19} and  fermionic superconducting system \cite{tul18} are important for nuclear physics too. The entanglement of valance particles is studied in the traditional nuclear shell model \cite{leg15}.

The aim of the present work is to investigate the mode entanglement in nuclear systems  in the simplest case i.e. in the case of 
two interacting particles described by a pure state assuming particle number conservation.
In the works \cite{kwa14,kwa16,kwa17} similar systems were considered and the von Neumann entropy of the one-particle reduced density operator of angular momentum and isospin coupled states were studied. Our present work can be considered as an extension and continuation of the research \cite{kwa14,kwa16,kwa17}. First of all, we study the mode entanglement instead of the particle entanglement and give 
explicit analytical expressions for measures of the entanglement and correlation. 
We also consider the interaction between the particles and use the configuration interaction method to describe ground and excited states with the interaction USD \cite{usd} and USDB \cite{usdb}.

This work is organized as follows. The formalism of the mode entanglement and two-mode correlation is briefly reviewed in section \ref{ent1}. The Slater decomposed form of a two-fermion wave function and the calculation of the entropies using this form of the wave function are discussed in section \ref{twof}. The angular momentum coupled single configurations and the states of the  configuration interaction method  are analysed in sections \ref{single} and \ref{ci}. The numerical results are shown in section \ref{num} where two neutron problem is considered in the $sd$ shell. Section \ref{sum} summarizes the results.
 
\section{Mode entanglement and two-mode correlation}\label{ent1}
Here we summarize the basic notions of the mode entanglement for a fermionic system.
This approach uses the language of the second quantized formalism of the non-relativistic quantum mechanics. We have a finite-dimensional single-particle (sp) Hilbert space $\cal S$ and from this, we construct the Fock space. If we take an 
orthonormal  basis $\{\phi_i\ \vert\ i=1,2,\ldots,d\} $ in $\cal S$ then 
the corresponding creation and annihilation operators acting on the Fock space are 
denoted by $c_i^\dagger$ and $c_i$, respectively.
We consider the canonical anti-commutation relation (CAR) algebra $\cal A$  generated by the set of operators $\{c_i, c_i^\dagger\ \vert\ i=1,\ldots,d\}$  whose elements satisfy the CARs
\begin{eqnarray}
c_ic_j^\dagger+c_j^\dagger c_i=\delta_{i,j},\nonumber\\
c_ic_j+c_jc_i=c^\dagger_ic_j^\dagger+c^\dagger_jc_i^\dagger=0,
\end{eqnarray}
where $i,j=1,\ldots,d$. The vacuum of the operators $c_i$ is denoted by  $\vert 0\rangle$
and we say that the system is described by $d$ fermionic modes.

Here we discuss the entanglement associated with  bipartitions of the modes. 
A bipartition for fermionic  system is defined by two subsets of modes and this will determine 
a bipartition of the algebra $\cal A$ in terms of subalgebras. First we define two sets of modes $A=\{\pi_1,\pi_2,\ldots,\pi_{a}\}$ and 
$B=\{\pi_{a+1},\ldots,\pi_d\}$, where $\pi_1,\pi_2,\ldots,\pi_d$ is a 
permutation of the numbers $1,2,\ldots,d$ and $1\leq a\leq d-1$. We will denote by ${\cal A}_A$ and ${\cal A}_B$ the operator subalgebras spanned by 
the modes $A$ and $B$, respectively.

First we consider such a bipartition where the subalgebra ${\cal A}_A$ is generated by the operators $c_k,c^\dagger_k$, i.e. the  single mode $k$. The CAR subalgebra 
${\cal A}_B$
is generated by the remaining modes. The one-mode reduced density matrix (OM-RDM) can be given in the form
\begin{equation}\label{onemden}
 \rho^{(k)}=\left(\begin{array}{cc}
\langle c_k^\dagger c_k\rangle&0\\
0& 1-\langle c_k^\dagger c_k\rangle
\end{array}\right),
\end{equation}
where the notation $\langle c_k^\dagger c_k\rangle=\langle\Psi\vert c_k^\dagger c_k\vert\Psi\rangle$ is used.
The pure state $\Psi$ describes the system. The one-mode reduced density operator $\hat\rho_A$ on the basis 
$\vert b_1\rangle=c_k^\dagger\vert 0\rangle$ and $\vert b_2\rangle= \vert 0\rangle$ is given  by
\begin{equation}
\hat\rho_A=\sum_{i,j=1}^2\vert b_i\rangle{ \rho}^{(k)}_{i,j}\langle b_j\vert.
\end{equation}
The reduced density operator $\hat \rho_A$ has the following important property \cite{gig15} $\Tr(\hat\rho_A   \hat O_A)=\langle\Psi\vert  \hat O_A\vert\Psi\rangle$, where $  \hat O_A$ is an arbitrary operator from ${\cal A}_A$.

The one-mode entropy is  given by
\begin{equation}\label{onemdef}
S(\rho^{(k)})=-\Tr( \rho^{(k)}\log_2( \rho^{(k)}))=h(\langle c_k^\dagger c_k\rangle),
\end{equation}
where the function $h$ is defined by $h(x)=-x\log_2(x)-(1-x)\log_2(1-x)$. The total correlation is the sum of the one-mode entropies\cite{leg04} 
\begin{equation}\label{totcor}
S_c=\sum_{i=1}^d S( \rho^{(i)}).
\end{equation}
This quantity depends on the choice of the sp basis\cite{kru15}.
We can consider the minimum of (\ref{totcor}) over all sp basis of $\cal S$ and define the one-body entanglement entropy \cite{gig15}
\begin{equation}
S^{\rm SP}={\rm min}_{c} S_{c}.
\end{equation}
The function $S^{SP}$ takes its minimum value zero for non-entangled states \cite{gig15}.
We have kept the naming of this entanglement measure and the symbol $S^{SP}$ introduced in the work \cite{gig15}.

A standard notion in many-body quantum physics is the one-particle reduced density matrix (OP-RDM)
\begin{equation}
 \rho^{sp}_{i,j}=\langle\Psi\vert c_j^\dagger c_i\vert\Psi\rangle.
\end{equation}
It was shown in \cite{gig15} that the one-body entanglement entropy $S^{SP}$ can be calculated using such a sp basis where the OP-RDM is diagonal $ \rho^{sp}={\rm diag}\{n_1,n_2,\ldots,n_d\}$ and
\begin{equation}\label{onespent}
S^{\rm SP}=\sum_{i=1}^d h(n_i).
\end{equation}
The occupation numbers  $n_i$ of the natural orbits are the eigenvalues of the OP-RDM.

Next we consider such a bipartition of the modes where two modes determine ${\cal A}_A$;
${\cal A}_A$ is generated by a pair of modes  
$(i;j)$
i.e. the operators $\{c_i,c_i^\dagger,c_j,c_j^\dagger\}$ ($i\neq j$). 
The two-mode reduced density matrix can be constructed using transition operators\cite{ris06,bar15,ser17}
which in the current case simplifies to \cite{tul18}
\begin{equation}\label{twomden}
 \rho^{(i;j)}=\left(\begin{array}{cccc}
\langle c_i^\dagger c_i c_j^\dagger c_j\rangle&0&0&0\\
0&\langle c_i^\dagger c_i c_j c^\dagger_j\rangle&\langle c_j^\dagger c_i\rangle&0\\
0&\langle c_i^\dagger c_j\rangle&\langle c_i c^\dagger_i c_j^\dagger c_j\rangle&0\\
0&0&0&\langle c_i c_i^\dagger c_j c^\dagger_j\rangle\\
\end{array}\right).
\end{equation}
The two-mode entropy of the modes $i$ and $j$ $S( \rho^{(i;j)})$ is the traditional von Neumann entropy of (\ref{twomden}) i.e. $-\Tr[ \rho^{(i;j)}\log_2( \rho^{(i;j)})]$.
It is easy to show that $S( \rho^{(i;j)})=S( \rho^{(j;i)})$. The two-mode reduced density operator $\hat\rho_A$ on the basis $\vert b_1\rangle=c_i^\dagger c_j^\dagger\vert 0\rangle$, $\vert b_2\rangle=c_i^\dagger\vert 0\rangle$, $\vert b_3\rangle=c_j^\dagger\vert 0\rangle$ and  $\vert b_4\rangle=\vert 0\rangle$ is given by
\begin{equation}
\hat\rho_A=\sum_{k,l=1}^4\vert b_k\rangle{ \rho}^{(i;j)}_{k,l}\langle b_l\vert.
\end{equation}
The reduced density operator $\hat\rho_A$ can be used to determine all expectation values of operators  $\hat O_A$ belong to ${\cal A}_A $ i.e.  $\Tr(\hat\rho_A   \hat O_A)=\langle\Psi\vert  \hat O_A\vert\Psi\rangle$.

The mutual information between the modes $i$ and $j$ ($i\neq j$) is defined in the following way
\begin{equation}\label{minfo}
{\cal I}^{(i;j)}=S( \rho^{(i)})+S( \rho^{(j)})-S( \rho^{(i;j)}).
\end{equation}
It describes the correlation between the modes $i$ and $j$ embedded into the environment of the other modes,
and it includes correlations of both classical and quantum origin~\cite{din20b}. 
\section{Two-fermion wave function}\label{twof}

We consider two identical fermions in a pure state $\vert\psi\rangle$.
The orthonormal one-fermion set of states is $\{\phi_i=c^\dagger_i\vert0\rangle, i=1,\ldots,d\}$.
The wave function $\vert\psi\rangle$ of a two-fermion state can be represented as 
\begin{equation}\label{wf}
\vert\psi\rangle=\sum_{i,j=1}^d w_{i,j} c_i^\dagger c_j^\dagger\vert 0\rangle,
\end{equation}
where the complex (or real) coefficients satisfy $w_{i,j}=-w_{j,i}$. The coefficients $w_{i,j}$ determine a 
skew--symmetric matrix $w$ and the  normalization condition is $2{\rm Tr}(ww^\dagger)=1$.  
\subsection{Slater decomposition}
There is a  classical theorem about skew-symmetric matrices which we will use in the following. 
A number of proofs can be found in the literature \cite{mat1,mat2,mat3,zum62,fiz2}.

{\it Theorem 1.} 
If $w$ is an even-dimensional complex (or real) non-singular $2n \times  2n$ skew-
symmetric matrix, then there exists a unitary (or real orthogonal) $2n \times 2n$ matrix $U$ such that:
\begin{eqnarray}\label{realform}
\fl&U^twU\nonumber\\
&={\rm diag}\left\{\left(\begin{array}{cc}
0&{\lambda_1}\\
-{\lambda_1}&0
\end{array}\right ),\left(\begin{array}{cc}
0&{\lambda_2}\\
-{\lambda_2}&0
\end{array}\right),\cdots,\left(\begin{array}{cc}
0&{\lambda_n}\\
-{\lambda_n}&0
\end{array}\right )
\right\}.
\end{eqnarray}
The rhs of (\ref{realform}) is written in block diagonal form with $2\times 2$ matrices appearing along the
diagonal, and the $\lambda_j$ are real and positive.

If $w$ is a complex (or real) singular skew-symmetric $d \times d$ matrix of rank $2n$ ($d > 2n$), then there exists a unitary (or real orthogonal) $d \times d$
matrix $U$ such that
\begin{eqnarray}\label{rform}
\fl&U^twU\nonumber\\
\fl&={\rm diag}\left\{\left(\begin{array}{cc}
0&{\lambda_1}\\
-{\lambda_1}&0
\end{array}\right ),\left(\begin{array}{cc}
0&{\lambda_2}\\
-{\lambda_2}&0
\end{array}\right),\cdots,\left(\begin{array}{cc}
0&{\lambda_n}\\
-{\lambda_n}&0
\end{array}\right ),{\cal O}_{d-2n}.
\right\}
\end{eqnarray}

The rhs of (\ref{rform}) is written in block diagonal form with $2\times 2$ matrices appearing along the
diagonal followed by a $(d-2n)\times (d-2n) $ block of zeros (denoted by ${\cal O}_{d-2n})$ and the
$\lambda_j$ are real and positive.

We use the so called canonical real form, where 
$\lambda_i$
are real and positive. 
The Slater decomposition given in \cite{pas01,sch01} is in terms of complex $\lambda_i$ but for our purposes it is worthwhile to choose $U$ 
so that we get the canonical real form \cite{fiz2,lev05}. In this case the quantities $\lambda_i$ are the 
non-zero singular values of $w$ i.e.
$\lambda^2_i$ are the non-zero real eigenvalues of the self-adjoint matrix $w^\dagger w$. These eigenvalues are double degenerated.

With the unitary transformation of the Theorem 1 we can define new creation and annihilation operators 
\begin{equation}\label{newm}
a_i^\dagger=\sum_{k=1}^{d}U^\dagger_{i,k} c_k^\dagger\quad{\rm and}\quad a_i=\sum_{k=1}^dU_{k,i} c_k.
\end{equation}
These new operators determine $d$ new modes\cite{kru15}. We can rewrite the wave function (\ref{wf}) into the  so called Slater decomposed form \cite{low56,ghi04}
\begin{equation}\label{wf2}
\vert\Psi\rangle=2\sum_{i=1}^{n} \lambda_{i}a_{2i-1}^\dagger a_{2i}^\dagger\vert0\rangle.
\end{equation}
We will call the new modes $(2i-1)$ and $2i$ as associated modes ($i=1,\ldots,n$).
The Slater decomposition introduces three sets  of the new modes $M_1=\{1,3,\ldots, 2n-1\}$, $M_2=\{2,4,\ldots, 2n\}$ and
$M_3=\{2n+1,2n+2,\ldots, d\}$. The modes in $M_3$ are not present in the state (\ref{wf2}).
Each mode from the sets $M_1$ or $M_2$ appears just in one component of (\ref{wf2}).
Furthermore Slater determinants constructed only from modes in $M_1$
($M_2$) are not present in (\ref{wf2}).
  
The equation (\ref{wf2}) is the fermionic analogue of the Schmidt decomposition  for distinguishable bipartite systems \cite{nie00,hor09}. It is customary to say that the Slater rank  of the wave function (\ref{wf}) 
is $n$.  The relation of Slater rank with the entanglement has been investigated in several papers \cite{sch01,eck02,maj16}. The state (\ref{wf}) is called entangled if and only if its Slater rank is greater than 1.

\subsection{Correlation measures using Slater decomposed form}\label{slatercor}

A two-particle state is characterized by a skew-symmetric matrix $w$. 
The quantities which measure the correlation can be expressed by $w$. For example the OP-RDM reads
\begin{equation}\label{rhow}
 \rho^{sp}=4 ww^\dagger.
\end{equation}
and the OM-RDM can be written in the form
\begin{equation}\label{onemdenrho}
 \rho^{(k)}=\left(\begin{array}{cc}
 \rho^{sp}_{k,k} &0\\
0& 1- \rho^{sp}_{k,k}\\
\end{array}\right).
\end{equation}

With a straightforward calculation  we can get an expression for the
two-mode density matrix (\ref{twomden})
\begin{equation}\label{twomden2}\fl
 \rho^{(i;j)}=\left(\begin{array}{cccc}
4\vert w_{i,j}|^2&0&0&0\\
0&-4\vert w_{i,j}|^2+ \rho^{sp}_{i,i}& \rho^{sp}_{i,j}&0\\
0& \rho^{sp}_{j,i}&-4\vert w_{i,j}|^2+ \rho^{sp}_{j,j}&0\\
0&0&0&1+4\vert w_{i,j}|^2- \rho^{sp}_{i,i}- \rho^{sp}_{j,j}\\
\end{array}\right).
\end{equation}

If the wave function is in Slater decomposed form we can get simpler expressions for the entanglement entropies.
Here we give the correlation measures expressed by the singular values $\lambda_i$. Let's assume that a pure state is given in the Slater decomposed form (\ref{wf2}).  The $w$ matrix of the wave function (\ref{wf2}) according to (\ref{realform}) and (\ref{rform}) is block diagonal $w={\rm diag}\left(\bar w,{\cal O}_{d-2n}\right)$ and $\bar w$ is $2n\times 2n$ type matrix with elements
\begin{eqnarray}\label{barw}
&\bar w_{i,j}=\delta_{|i-j|,1}\left\{
\begin{array}{ll}
\lambda_{\lceil i/2\rceil }\delta_{(-1)^i,-1}&{\rm if}\quad i<j\\
-\lambda_{\lceil i/2\rceil}\delta_{(-1)^i,1}&{\rm if }\quad i>j
\end{array}
\right.\nonumber\\
&i,j=1,\ldots,2n,
\end{eqnarray}
where $\lceil x\rceil$ is the ceiling function.
From (\ref{rhow}) and (\ref{barw}) it follows that the OP-RDM is also  block diagonal $\rho^{sp}={\rm diag}\left(\bar\rho^{sp},{\cal O}_{d-2n}\right)$ and $\bar\rho^{sp}$ is a diagonal $2n\times 2n$ type  matrix
\begin{equation}\label{onepden}
\bar\rho^{sp}_{i,j}=
\delta_{i,j}4\lambda^2_{\lceil i/2\rceil} ,\quad i,j=1,\ldots,2n.
\end{equation}

The one-mode entropies of the modes  for the state (\ref{wf2}) can be calculated using 
(\ref{onemdef}), (\ref{onemdenrho}) and (\ref{onepden})
\begin{equation}\label{ome}
S( \rho^{(2i-1)})=S( \rho^{(2i)})=h(4 \lambda_{i}^2) , \ \ \ i=1,\ldots,n.
\end{equation}
Notice that the associated modes have the same one-mode entropies. The one-mode entropies of the remaining $d-2n$ modes are zero $S(\rho^{(i)})=0,\ \ i=2n+1,\ldots d$.
The one-body entanglement entropy according to (\ref{onespent}) and (\ref{ome}) is
\begin{equation}
S^{SP}=\sum_{i=1}^{2n}h\left(4 \lambda_{\lceil i/2\rceil}^2\right)=2 \sum_{i=1}^n h\left(4 \lambda_{i}^2\right).
\end{equation}
We might ask the question which two-body state has the maximal one-body entanglement entropy if we can use only  predefined number of modes.
In the Appendix we show that the maximum value of $S^{SP}$ is 
\begin{equation}\label{sspmax}
-2+d \log_2 d-(d-2)\log_2(d-2)
\end{equation}
if  $d>2$ and in this case $\lambda_i^2=\frac{1}{2d}$. If $d=2$ the state is non-entangled and $S^{SP}=0$.

Considering (\ref{twomden2}), (\ref{barw}) and (\ref{onepden}) we can realize that the two-mode reduced density matrix is a  diagonal $4\times4$ matrix in a sp basis which 
corresponds to Slater decomposed form. For non associated modes we can write  
\begin{equation}\label{rhoij1}
 \rho^{(i;j)}=\rm diag\{0,4\lambda_{\lceil i/2\rceil}^2,4\lambda_{\lceil j/2\rceil }^2,1-4\lambda_{\lceil i/2
\rceil }^2-4\lambda_{\lceil j/2\rceil}^2\}, 
\end{equation}
where $i,j=1,\ldots,2n$ and $i\neq j$.
The case when the two modes are associated is interesting.  
If  $i$ is odd then we have 
\begin{equation}\label{rhoij2}
 \rho^{(i;i+1)}={\rm diag}(4\lambda_{\lceil i/2\rceil}^2,0,0,1-4\lambda_{\lceil i/2\rceil}^2),\ \ \ i=1,3,\ldots,2n-1.
\end{equation}
This means that according to (\ref{ome}) for associated modes we have the relations
\begin{equation}\label{rhoass}
S( \rho^{(2i-1)})=S( \rho^{(2i)})=S( \rho^{(2i-1;2i)})=h(4\lambda_i^2),\ \ \ i=1,2,\ldots,n
\end{equation}
i.e. the one-mode entropies of the associated modes are the same, furthermore this value agrees with the two-mode entropy of the modes $(2i-1)$ and $2i$. 
From these properties it follows that the mutual information (\ref{minfo}) of the associated modes  is the same as the one-mode entanglement entropy of the considered modes.
For the remaining $d-2n$ modes we have $S(\rho^{(i;j)})=0$ if $2n+1\leq i\leq d$ or $2n+1\leq j\leq d$ and $i\neq j$.
\section{States with single configuration}\label{single}
In the rest of the paper, we consider an interacting system composed of identical fermions (either protons or neutrons) where the Hamiltonian is spherical symmetric. In this case, the square of the total angular momentum is a conserved quantity. We have to construct wave function with good total angular momentum.
In this section, we apply the general formalism developed in the previous sections for the study of the entanglement ``caused'' by angular momentum coupling. Of course, the real reason of the entanglement is the spherical symmetric interaction between the fermions.

We use the following notation for the modes (sp states) $\vert\phi_\alpha\rangle=\vert\alpha\rangle=\vert a m_\alpha\rangle$  
where $a=n_a l_a j_a$. Here $l_a$ and $j_a$ have the meanings as the quantum numbers for orbital
and total angular momenta of the sp orbit $a$, and $n_a$ is as an additional
quantum number to  fully characterize the mode. In the followings a mode will be denoted by $(\alpha)=(a,m)$.  The angular momentum coupling of the sp orbits is signed by 
$[a_1\otimes a_2]^{JM}$ and the explicit form of 
a two nucleon wave function with  total angular momentum $J$ and projection $M$ is  
\begin{equation}\label{j2wf}
\vert\Phi^{JM}_{a_1a_2}\rangle=N_{a_1,a_2}\sum_{m_1,m_2}\langle j_{a_1},m_1,j_{a_2},m_2\vert J,M\rangle  c_{a_1m_1}^\dagger c_{a_2m_2}^\dagger\vert 0\rangle.
\end{equation}
The normalization factor is $N_{a_1,a_2}=1$ if $a_1\neq a_2$, and for  $a_1=a_2$ we have 
$N_{a_1,a_2}=1/\sqrt{2}$ (in this case $J$ is an even integer).  We denote the number of modes which determines the sp space by $d$. It is obvious that $d=2j_{a_1}+2j_{a_2}+2$ if $a_1\neq a_2$ and $d=2j_a+1$ if $a_1=a_2=a$. 
The size of the matrix $w$, which corresponds to the state (\ref{j2wf}) is $d$.

\subsection{Slater decomposition}\label{sdecomp}

Here we illustrate that we can arrange the modes in such a way that the state  (\ref{j2wf}) is in  Slater decomposed form, and we can avoid the numerical calculation of the singular values of the matrix $w$.  In this way we can give analytical expressions for the entropies.

Pairs of modes which satisfy the condition $M=m_1+m_2$ are arranged in the following order $[(a_1,\bar m),(a_2,M-\bar m)],[(a_1,\bar m +1),(a_2,M-\bar m-1)],\ldots,[(a_1,\bar M),(a_2,M-\bar  M)]$. We call  this arrangement as mode sequence $\cal M$. The lower and upper limits are defined by
\begin{equation}\label{meq}
\bar m=\max(-j_{a_2}+M,-j_{a_1})
\end{equation}
and
\begin{equation}\label{Meq}
\bar M=\min(j_{a_1},j_{a_2}+M).
\end{equation}
In order to make the equations more readable we do not denote that $\bar m$ and $\bar M$ depend on the quantum numbers $j_{a_1}$, $j_{a_2}$ and $M$.
The square brackets $[\ldots,\ldots]$ denote pairs of modes which 
determine two-particle states of the form:
$c^\dagger_{a_1\bar m}c^\dagger_{a_2(M-\bar m)}\vert0\rangle,\ldots,c^\dagger_{a_1\bar M}c^\dagger_{a_2(M-\bar M)}\vert0\rangle$.
In the case of  $a_1=a_2$ there are identical pairs of modes in the sequence $\cal M$. 
We remove pairs of modes from the end of $\cal M$ in such a way that each mode pair should be in the sequence $\cal M$ only once.

In order to have  canonical real form of the state (\ref{j2wf}) as described in Theorem 1 we have to make  modifications in sequence $\cal M$.  If the Clebsch-Gordan coefficient $\langle j_{a_1},m,j_{a_2},M-m\vert JM\rangle$ of a mode pair is negative we have to interchange the order of modes in the pair. In some cases we have to make a further modification of $\cal M$. If a pair of mode has the same quantum numbers $\alpha$ then the corresponding two particle wave function is identically zero ($c_\alpha^\dagger c_\alpha^\dagger\vert0\rangle=0$) therefore we remove the mode pair $[\alpha,\alpha]$ from the sequence $\cal M$  and put the mode $(\alpha)$ 
into a set called $\cal N$.  The modes which are not contained in $\cal M$ are collected in $\cal N$.

Finally we have to consider a peculiarity of Clebsch-Gordan coefficients. There are so called exceptional cases where the quantum numbers satisfy all obvious symmetries e.g. triangular inequality etc., nevertheless the value of the Clebsch-Gordan coefficient is zero \cite{ray78, hei09}. 
We have to leave out those pairs of modes from $\cal M$  whose Clebsch-Gordan coefficients 
$\langle j_{a_1},m,j_{a_2},M-m\vert JM\rangle$ are exceptional ones and move these modes into the set $\cal N$. The number of exceptional Clebsch-Gordan coefficients is denoted by $N_{exc}$. The number of modes which remains in the mode sequence $\cal M$ 
is denoted by $2n$ in accordance with the previous sections and
\begin{equation}\label{n2}
2n=\left\{\begin{array}{ll}
2\lfloor(\bar M-\bar m+1)/2\rfloor-2N_{exc}&\quad{\rm if}\quad a_1=a_2\nonumber\\
2(\bar M-\bar m+1)-2N_{exc}&\quad{\rm if}\quad a_1\neq a_2,\\
\end{array}\right.
\end{equation}
where the floor function is signed by $\lfloor x\rfloor$.
The pairs of modes remained in the sequence $\cal M$ are called associated modes and they are of the form  $(a_1,m)$ and $(a_2,M-m)$ ($\bar m\leq m\leq \bar M$).

Once we have the final sequence of modes $\cal M$,  we  give a unique serial number for each modes in $\cal M$ and define the mapping  
$\alpha(i)=(a(i),m(i))=(n(i),l(i),j(i),m(i))$ $i=1,2,\ldots,2n$ in such a way that it corresponds to the final mode sequence $\cal M$. The notation $\alpha(i)$ means the quantum numbers  of the $i$'th mode in the ordered  enumeration $\cal M$. We can say the number of modes present in the wave function (\ref{j2wf}) is $2n$.
 
With these notation we can rewrite (\ref{j2wf}) in Slater decomposed form
\begin{equation}\label{sj1j2}
\vert\Phi_{a_1a_2}^{JM}\rangle=\sum_{i=1}^{n}2 \lambda_i\ c_{\alpha(2i-1)}^\dagger c_{\alpha(2i)} ^\dagger\vert 0\rangle,
\end{equation}
where the $\lambda_i$ is expressed by Clebsch-Gordan coefficients
\begin{eqnarray}\label{lambda}\fl
&\lambda_i=\left\{
\begin{array}{ll}
\frac{1}{\sqrt{2}}
\left\langle j(2i-1),m(2i-1) ,j(2i),m(2i)\vert JM\right\rangle\quad &{\rm if}\quad a_1=a_2\\
\frac{1}{2}
\left\vert\left\langle j(2i-1),m(2i-1) ,j(2i),m(2i)\vert JM\right\rangle\right\vert\quad &  {\rm if}\quad  a_1\neq a_2,\nonumber\\
\end{array}\right.\\
&i=1,2,\ldots,n.
\end{eqnarray}

\subsection{Entanglement entropies}\label{nodyn}
Since we have rewritten the angular momentum coupled two-fermion states in Slater decomposed form we can easily get the entanglement entropies  from the general formulas of section \ref{slatercor}. 

From (\ref{sj1j2}) it follows that the Slater rank is $S_{\rm rank}(a_1,a_2,M)=n$.
The following relation is valid
\begin{equation}\label{sr1}
S_{\rm rank}(a_1,a_2,M)=S_{\rm rank}(a_1,a_2,-M)
\end{equation}
and if we assume that  $N_{exc}=0$ both for $M$ and $M+1$
\begin{equation}\label{sr2}
S_{\rm rank}(a_1,a_2,M+1)\leq S_{\rm rank}(a_1,a_2,M),\quad 0\leq M\leq J-1.
\end{equation}
The smallest Slater rank corresponds 
to the cases $\vert M \vert=J$ if $N_{exc}=0$ for each $M$ ($-J\leq M\leq J$). These formulas are proved in the Appendix B.

From now on instead of the serial number of the modes we will identify them by their quantum numbers.
The one-mode entropies of the associated modes are identical and they are given by 
\begin{eqnarray}\label{j1j2onem}\fl
&S\left( \rho^{(a_1,m)}\right)=S\left( \rho^{(a_2,M-m)}\right)\nonumber\\
&=\left\{
\begin{array}{ll}
h\left(2\vert\left\langle j_{a_1}, m,j_{a_2},M-m\vert J,M\right\rangle\vert^2
\right)&\quad{\rm if}\quad a_1=a_2\\
h\left(\vert\left\langle j_{a_1}, m,j_{a_2},M-m\vert J,M\right\rangle\vert^2
\right)&\quad{\rm if}\quad a_1\neq a_2\\
\end{array}\right.\\
&\bar m\leq m\leq\bar M\nonumber,
\end{eqnarray}
where we used (\ref{ome}) and (\ref{lambda}). The one-mode entropies of the remaining  modes are zero.
The two-mode entropies can be easily calculated with the help of (\ref{rhoij1}), (\ref{rhoij2}) and (\ref{lambda}).

In the very important case when $J=0$ the expressions above have simpler forms. In this case we always have $j_{a_1}=j_{a_2}=j$. Fortunately there are no exceptional Clebsch-Gordan coefficients in the case  i.e.  $N_{exc}=0$ in (\ref{n2}). This fact helps to derive simple analytical expressions. The Slater rank is
\begin{equation}\label{srank}
S_{\rm rank}(a_1,a_2,0)=\left\{\begin{array}{ll}
(2 j+1)/2&\quad{\rm if}\quad a_1=a_2\\
(2j+1)&\quad{\rm if}\quad a_1\neq a_2
\end{array}\right..
\end{equation}

From a property of the Clebsch-Gordan coefficients, $\langle j,m,j,-m\vert 0,0\rangle=(-1)^{j-m}/\sqrt{2j+1}$ and from ({\ref{sj1j2}) and ({\ref{lambda}) it follows that 
\begin{eqnarray}\label{lam4}\fl
&\langle\Phi^{00}_{a_1a_2}\vert c_{\alpha(i)}^\dagger c_{\alpha(i)}\vert\Phi^{00}_{a_1a_2}\rangle=4\lambda_{\lceil i/2\rceil}^2=\left\{\begin{array}{ll}
2/(2j+1)&\quad{\rm if}\quad a_1=a_2\\
1/(2j+1)&\quad{\rm if}\quad a_1\neq a_2
\end{array}
\right.\nonumber\\
&i=1,2,\ldots,2n.
\end{eqnarray}
According to (\ref{onepden}) and (\ref{lam4}) the diagonal elements of the diagonal OP-RDM are identical.
From this observation and from (\ref{ome}) we get for the state $\vert\Phi^{00}_{a_1a_2}\rangle$ that
the one-mode entropies of  each modes are identical i.e. they do not  depend on $m$ 
\begin{eqnarray}\label{onemj0}
S\left( \rho^{(a,m)}\right)=
h\left(\frac{2}{2j+1}\right),\quad 
a_1=a_2=a,\quad
m=-j,\ldots,j
\end{eqnarray} 
and
\begin{eqnarray}\label{onemjj0}\fl
S\left( \rho^{(a_1,m)}\right)=S\left( \rho^{(a_2,m)}\right)=
h\left(\frac{1}{2j+1}\right),\quad 
a_1\neq a_2,\quad
m=-j,\ldots,j.
\end{eqnarray} 
In the case of $J=0$ the one-body entanglement entropy (\ref{onespent}) is 
\begin{equation}\label{j0onem}
S^{SP}=\left\{\begin{array}{ll}
(2j+1) h\left(\frac{2}{2j+1}\right)&\quad{\rm if}\quad a_1=a_2\\
2(2j+1) h\left(\frac{1}{2j+1}\right)&\quad{\rm if}\quad a_1\neq a_2.
\end{array}
\right.
\end{equation}

The one-body entanglement entropy of two-nucleon states with total angular momentum zero shows an interesting property. The state $\vert\Phi_{aa}^{00}\rangle$  when $j=\frac{1}{2}$ is not entangled since its Slater rank is one. 
We might ask which two-body state has the maximal one-body entanglement entropy if we can use only the modes $(a,m)$ ($-j\leq m\leq j$). 
The states $\vert \Phi_{aa}^{00}\rangle$  ($j> \frac{1}{2}$) are maximally entangled in the previous sense.
The  OP-RDM of the state $\vert\Phi_{aa}^{00}\rangle$ is diagonal and the diagonal elements are identical and  their values are $2/(2j+1)$. The number of modes in $\Phi_{aa}^{00}$ is $(2j+1)$ and according to Appendix A this means that the state is maximally entangled.

In the case $J=0$ the two-mode entropies also do not depend on the $z$ components of the angular momentum  of the modes. For non associated modes from (\ref{rhoij1}) and (\ref{lam4}) we get for the two-mode entropies
\begin{eqnarray}\fl
&S\left( \rho^{((a_1,m_1);(a_2,m_2))}\right)=\left\{\begin{array}{ll}
2H\left(\frac{2}{2j+1}\right)+H\left(1-\frac{4}{2j+1}\right)&\ \ {\rm if}\ \  a_1=a_2,  m_1\neq \pm m_2\\
2H\left(\frac{1}{2j+1}\right)+H\left(1-\frac{2}{2j+1}\right)&\ \ {\rm if}\ \  a_1\neq a_2,  m_1\neq -m_2
\end{array}\right.\nonumber\\
& m_1,m_2=-j,-j+1,\ldots,j,
\end{eqnarray}
where $H(x)=-x\log_2(x)$.
For the associated modes we can use the general relation (\ref{rhoass}) and equations (\ref{onemj0}) and  (\ref{onemjj0}).

\section{\label{ci}States with configuration mixing}
In realistic calculations, the configurations are mixed and very frequently the  so-called CI method is used to solve the Schr\"odinger equation.
In this section, we explicitly give the mode entanglement measures when the wave function is in a special CI form and the total angular momentum is zero.
\subsection{\label{ci2} Configuration interaction method}
The general trial wave function in the CI method is of the form 
\begin{equation}\label{ciwfg}
\vert\Psi^{(J=0,M=0)}\rangle=\sum_{a_1,a_2} A_{a_1,a_2}\vert\Phi_{a_1a_2}^{00}\rangle,
\end{equation}
where the amplitudes $A_{a_1,a_2}$ are determined by the Rayleigh-Ritz variational principle and $j_{a_1}=j_{a_2}$.
If the self-adjoint Hamiltonian is invariant with respect to the complex conjugation then $A_{a_1,a_2}$ can be chosen to be real. 
It is easy to see that the wave function (\ref{ciwfg}) generally cannot be  transformed into Slater decomposed form with the help of special order of the modes. However the following form of CI wave function 
\begin{equation}\label{ciwf}
\vert\Psi^{(J=0,M=0)}\rangle=\sum_{a} A_{a}\vert\Phi_{aa}^{00}\rangle
\end{equation}
can be brought to Slater decomposed form with a special order of the modes.
Assume that the possible orbits are $a_1,a_2,\ldots a_k$ in the summation (\ref{ciwf}) and $A_a\neq 0$.
The modes are arranged in the form: modes of orbit $a_1$, modes of orbit  $a_2$ and so on. At each $a_i$ the modes $(a_i,m)$ are arranged in the form as described in the section \ref{sdecomp}. If $A_a$ is negative then in the mode sequence $\cal M$, which corresponds to the orbit $a$, the order of the modes in each mode pair has to be interchanged. It is obvious that the CI  wave function (\ref{ciwf}) with this mode arrangement is in Slater decomposed form. From (\ref{srank}) it follows that  the Slater rank of the CI wave function (\ref{ciwf}) is 
$
\frac{1}{2}\sum_a (2 j_a +1).
$
For the one-mode entropies according to (\ref{onemden}) we need the expression
\begin{equation}
\langle\Psi^{(0,0)}\vert c_{am}^\dagger c_{am}\vert\Psi^{(0,0)}\rangle=
\vert A_{a}\vert ^2\langle\Phi^{00}_{aa}\vert c_{am}^\dagger c_{am}\vert\Phi^{00}_{aa}\rangle.
\end{equation} 
From the equation above and from (\ref{onemdef}) and (\ref{lam4}) it follows that for the wave function (\ref{ciwf})  
the one-mode entropies of the modes $(a,-j),(a,-j+1),\ldots,(a,j-1),(a,j)$ are 
\begin{eqnarray}\label{cionem}
S( \rho^{(a,m)})=h\left(\vert A_{a}\vert ^2\frac{2}{2j_{a}+1}\right).
\end{eqnarray}
This expression shows that the one-mode entropies of the modes $(a,m)$ do not depend on the magnetic quantum number $m$. This was observed numerically in \cite{leg15} for systems of few fermions; here we proved it rigorously for two-fermion systems with CI wave function of the form (\ref{ciwf}).

The one-body entanglement entropy of the wave function (\ref{ciwf}) has the obvious form 
\begin{equation}
S^{SP}=\sum_{a} (2j_a+1)h\left(\vert A_a\vert ^2\frac{2}{2j_a+1}\right).
\end{equation} 

We can determine a CI wave function which is in the form (\ref{ciwf}) and  it has the largest one-body entanglement entropy. According to the Appendix A 
the occupation numbers of the natural orbits have to be the same 
and this value has to be two over the number of modes present in the wave function. We have the equation 
\begin{equation}\label{maxci}
\frac{2\vert A_a\vert^2}{2j_a+1}=\frac{2}{\sum_{a'} (2j_{a'}+1)}
\end{equation}
and from this equation we can get $\vert A_a\vert$. However we can not determine the sign of the amplitude $A_a$.
In the case of maximally entangled CI wave function the one-mode entropies are independent from the modes 
\begin{equation}\label{indentr}
S( \rho^{(a,m)})=h\left(\frac{2}{\sum_{a'} (2j_{a'}+1)}\right),\ \  m=-j,\ldots,j.
\end{equation}

The two-modes entropies can be expressed also in simple forms for the state (\ref{ciwf}).
It is obvious that the modes $(a_i,m_k)$ and 
$(a_j,m_l)$ are not associated if $a_i\neq a_j$. Using the general expressions (\ref{rhoij1}) and (\ref{lam4}) for
 two-mode entropies we get
\begin{eqnarray}\label{2m1}
S\left( \rho^{((a_i,m_k);(a_j,m_l))}\right)&=&H\left(\frac{2\vert A_{a_i}\vert^2}{2j_{a_i}+1}\right)+H\left(\frac{2\vert A_{a_j}\vert^2}{2j_{a_j}+1}\right)\nonumber\\
&+&H\left(1-\frac{2\vert A_{a_i}\vert ^2}{2j_{a_i}+1}-\frac{2\vert A_{a_j}\vert^2}{2j_{a_j}+1}\right),\quad
a_i\neq a_j.
\end{eqnarray}
In the same way we can get for two-mode entropies of the  non-associated modes $(a,m_1)$ and $(a,m_2)$ 
\begin{eqnarray}\label{2m2}
&S\left( \rho^{((a,m_1);(a,m_2))}\right)=2H\left(\frac{2\vert A_{a}\vert^2}{2j_a+1}\right)
+H\left(1-\frac{4\vert A_{a}\vert ^2}{2j_a+1}\right),\nonumber\\
&m_1,m_2=-j,\ldots,j,\ \ \ m_1\neq -m_2, \ \ m_1\neq m_2.
\end{eqnarray}
If the considered modes are associated i.e. the  modes are 
$(a,-m)$ and $(a,m)$ $(m=-j,-j+1,\ldots,j$) the two-mode entropy is 
\begin{equation}\label{2m3}\fl
S\left( \rho^{((a,-m);(a,m))}\right)=H\left(\frac{2\vert A_{a}\vert^2}{2j_a+1}\right)
+ H\left(1-\frac{2\vert A_{a}\vert^2}{2j_a+1}\right)=h\left(\frac{2\vert A_{a}\vert^2}{2j_a+1}\right).
\end{equation}
If we compare (\ref{2m3}) with (\ref{cionem}) we can realize that the one-mode entropies of the modes $(a,m)$ and $(a,-m)$ agrees with the two-mode entropies of the same modes i.e $S( \rho^{(a,-m)})=S\left( \rho^{(a,m)}\right)=
S\left( \rho^{((a,-m);(a,m))}\right)$.
The expressions (\ref{2m1}), (\ref{2m2}) and (\ref{2m3}) define three families of two-mode entropies but within 
one family the two-mode entropies do not depend on the magnetic quantum numbers.

\section{Numerical results}\label{num}

Here we consider two neutrons in the $sd$ shell and apply the CI method. The sp orbits are $1s_{1/2}$, $0d_{3/2}$ and $0d_{5/2}$ and 
the number of modes is twelve.  The ground and two excited $J^\pi=0^+$ states are determined with the interaction USD and USDB. In this case the CI wave function has the form (\ref{ciwf}) and we can apply for the entropies the analytical results of the section \ref{ci}. The energies are calculated with the CI method and shown in Table 1. The values of the amplitudes of the components of the wave functions are also displayed. The angular momentum coupled components are 
$[1s_{1/2}\otimes 1s_{1/2}]^{00}$,  $[0d_{3/2}\otimes 0d_{3/2}]^{00}$ and  $[0d_{5/2}\otimes 0d_{5/2}]^{00}$.
In each state the modulus of one component has much larger value than the modulus of the other components of the wave function. The dominant configurations are $(0d_{5/2})^2$,  $(1s_{1/2})^2$ and  $(0d_{3/2})^2$ for the states $0^+_1$, $0^+_2$ and $0^+_3$, respectively.

\begin{table}[h]
\caption{\label{energy}Energy of the two neutron system in the $sd$ shell using the CI method with the interaction USD. The ground and excited $J^\pi=0^+$ states are considered.  The  amplitudes of the decomposition into  angular momentum coupled configurations are displayed also.}
\begin{indented}
\item[]\begin{tabular}{|c|c|c|c|c|}
\br
state& E (MeV)&($s_{1/2})^2$&$(d_{3/2})^2$&$(d_{5/2})^2$\\
\br
$0^+_1$&-12.171&0.389&0.245&0.889\\
\hline
$0^+_2$&-7.851&0.919&-0.029&-0.393\\
\hline
$0^+_3$&1.964&0.071&-0.969&0.236\\
\br
\end{tabular}
\end{indented}
\end{table}

Table 2 contains one-body entanglement entropy together with the one-mode entropies. The ground state has 
the largest one-body entanglement entropy. 
To study the dependence of the one-body entanglement entropy on the amplitudes we take the following normalized ansatz
\begin{equation}\label{wave}\fl
\cos(\theta)\left[1s_{\frac{1}{2}}\otimes 1s_{\frac{1}{2}}\right]^{00}+\sin(\theta)\sin(\phi)\left[0d_{\frac{3}{2}}\otimes 0d_{\frac{3}{2}}\right]^{00}+\sin(\theta)\cos(\phi)\left[0d_{\frac{5}{2}}\otimes 0d_{\frac{5}{2}}\right]^{00}
\end{equation}
with real parameters $\theta$ and $\phi$. The results are depicted on Figure 1.
Since the  one-body entanglement entropy depends on only the moduli of the amplitudes we assume that both $\theta$ and $\phi$ are nonnegative and $\theta,\phi\leq \pi/2$. The small values of the entropy along the $\theta$ axis of the coordinate system is due to the fact that when $\theta=0$ the state (\ref{wave}) is non-entangled.
The maximum  value of $S^{SP}$ according to (\ref{sspmax}) is $7.80$ for such a CI wave function which has the same
components as we use in our CI calculations. In the maximum entropy case with the help of (\ref{indentr}) we can calculate the one-mode entropies  ($0.65$) they are independent from the modes.

\begin{table}\caption{One-mode and one-body entanglement entropies of the two-neutron $J^\pi=0^+$ states in the $sd$ shell using the CI method with the interaction USD and USDB. The one-mode entropy of the mode $(a,m)$ does not depend on $m$. }

\begin{indented}
\item[]\begin{tabular}{|c|c|c|c|}
\br
state&&CI USD&CI USDB\\
\hline
\multirow{5}{*}{$0^+_1$}&$S^{SP}$&6.988&6.951\\\cline{2-4}

&$s_{1/2}$&0.611&0.669\\\cline{2-4}
&$d_{3/2}$&0.194&0.166\\\cline{2-4}
&$d_{5/2}$&0.832&0.825\\
\hline\hline
\multirow{5}{*}{$0^+_2$}&$S^{SP}$&3.024&3.327\\\cline{2-4}
&$s_{1/2}$&0.623&0.676\\\cline{2-4}
&$d_{3/2}$&0.005&0.009\\\cline{2-4}
&$d_{5/2}$&0.293&0.323\\
\hline\hline
\multirow{5}{*}{$0^+_3$}&$S^{SP}$&4.881&4.759\\\cline{2-4}
&$s_{1/2}$&0.046&0.033\\\cline{2-4}
&$d_{3/2}$&0.997&0.998\\\cline{2-4}
&$d_{5/2}$&0.133&0.117\\\cline{2-4}
\br
\end{tabular}
\end{indented}
\end{table}

The maximum value of the $S^{SP}$ corresponds to such amplitudes  which satisfy  (\ref{maxci}). In our case we get the following absolute values for the  amplitudes $0.41$, $0.58$ and $0.71$ for the configurations  $(1s_{1/2})^2$,  $(0d_{3/2})^2$ and  $(0d_{5/2})^2$, respectively. For completeness we mention  that with the interaction  USD  the energy of the two-neutron system  is $-10.271$ MeV with the amplitudes  $0.41$, $0.58$ and $0.71$  for the configurations  $(1s_{1/2})^2$,  $(0d_{3/2})^2$ and  $(0d_{5/2})^2$.

The position of the maximum of the $S^{SP}$ is denoted  MAX in Figure 1.
The value of the $S^{SP}$ when the CI method used with interaction USD is signed by the abbreviation USD in Figure 1. One can notice that the one-body entanglement entropy of the ground state is close to the maximum value of the $S^{SP}$. We may say that interaction USD and USDB induce such a nucleon-nucleon correlation which leads to a ground state  wave function  which is almost maximally entangled with respect to the $S^{SP}$ measure.

We now discuss 
the relationship between the amplitudes and the one-mode entropies for the wave function (\ref{ciwf}). According to the expression  (\ref{cionem}) we have to know 
the characteristics of the function $h(x)$.  The function $h(x)$ on the interval $[0,1]$ has the following properties. The absolute maximum is at $x=1/2$ and here the one-mode entropy is one. At $x=0$ and $x=1$ the one-mode entropy is zero.
The function $h(x)$ is strictly increasing (decreasing) in the interval $[0,1/2]$($[1/2,1]$) and symmetric to the point $x=1/2$ i.e. $h(1/2-x)=h(1/2+x)$ ($h(x)=h(1-x)$). Since the argument of $h$ is $2\vert A_a\vert^2/(2j_a+1)$ we can make the following statements. If $\vert A_a\vert^2$ increases the one-mode entropy increases if $j_a>1/2$. The situation is different when $j_a=1/2$. In this case if $\vert A_a\vert^2$ increases the one-mode entropy increases too provided $\vert A_a\vert^2\leq 1/2$ is valid. If the modulus of the amplitude increases 
the one-mode entropy decreases if $\vert A_a\vert^2\geq 1/2$. These observations are in line with the physical expectations. 
At the first thought we may find strange the following. The moduli of amplitudes of the components $(s_{1/2})^2$ for the states $0_1^+$ and $0_2^+$  
are very different ($0.389$ and $0.919$). Nevertheless the one-mode entropies of the modes $(s_{1/2},m)$ are very similar ($0.611$ and $0.623$).
This seemingly strange result  follows from the symmetry property of the function $h(x)$.

\begin{figure}[h]\caption{\label{fig1}The one-body entanglement entropy as the function of the parameters
 $\theta$ and $\phi$ using the wave function (\ref{wave}).  The maximum value of the entropy  is denoted by MAX. The result of the CI calculation with interaction USD is signed by USD.}
 \vskip 0.2cm
\includegraphics[scale=0.6]{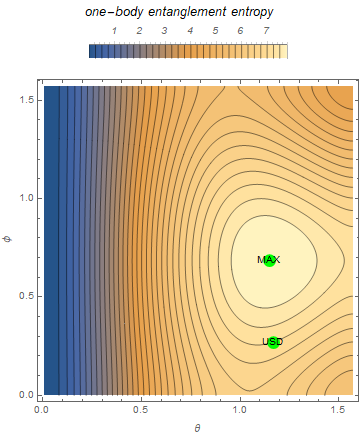}
\end{figure}

The mutual informations of two modes are displayed in Table 3. The largest mutual information corresponds to  such a pair of modes whose orbits belong to the wave function component whose amplitude has the largest modulus. It is interesting that the associated modes have much larger mutual information than the other pairs of modes. 
The tables presented in this section show that the USD and USDB interactions behave qualitatively in the same way.
\begin{table}
\caption{Mutual information of the modes.  The two-neutron ground state ($J^\pi=0^+$) in the $sd$ shell using the CI method with the interaction USD and USDB is considered. The wave function is in Slater decomposed form.
The sign (a) means that the considered modes are associated.}
\begin{indented}
\item[]\begin{tabular}{|l|c|c|}
\br
&CI USD&CI USDB\\
\hline
$(s_{1/2};s_{1/2})$ (a)&0.611&0.669\\
\hline
$(d_{3/2};d_{3/2})$ (a)&0.194&0.166\\
\hline
$(d_{5/2};d_{5/2})$ (a)&0.832&0.825\\
\hline
$(d_{3/2};d_{3/2})$ &0.001&0.001\\
\hline
$(d_{5/2};d_{5/2})$ &0.140&0.133\\
\hline
$(s_{1/2};d_{3/2})$&0.007&0.007\\
\hline
$(s_{1/2};d_{5/2})$&0.074&0.085\\
\hline
$(d_{3/2};d_{5/2})$&0.014&0.011\\
\br
\end{tabular}
\end{indented}
\end{table}

\section{Summary}\label{sum}

The mode entanglement and correlation are studied in the case of two interacting fermions. We showed that for single configurations properly choosing the order of the modes the angular momentum coupled wave function can be written in Slater decomposed form. 
The one- and two-mode entropies can be expressed with the help of the Clebsch-Gordan coefficients. If the fermions are coupled to total angular momentum zero then the expressions for the entanglement measures further simplified and very simple analytical forms are derived.  
When the two nucleons have zero total angular momentum in the configuration $(a)^2$ ($j_a\neq \frac{1}{2}$) then their states are maximally entangled with respect to the measure called  one-body entanglement entropy. 

In the case of configuration interaction method we considered a special form of wave function with total angular momentum zero.
In this case we also derived  simple formulas for  the calculation of the one- and two-mode entropy and determined the modulus of the amplitudes of the state which has maximum one-body entanglement entropy.

We carried out numerical calculations using the CI method in the case of two neutrons in the $sd$ shell using the interactions  USD and USDB.
We considered three states where the total angular momentum is zero. The ground state has the largest one-body entanglement entropy and this state is almost maximally entangled
with respect to the one-body entanglement entropy.  Using the Slater decomposition one can introduce the notion of associated modes. It turns out that for the ground state the mutual information of the associated modes
are much larger than the mutual information of any other pairs of modes.

\ack{This work has been supported by the National Research, Development and Innovation Fund of Hungary, financed under the K18
funding scheme with project nos. K 128729 and K 120569  and by the National Research, Development, and Innovation Office (NKFIH) 
through the Hungarian Quantum Technology National Excellence Program, project no. 2017-1.2.1- NKP-2017-0000. 
{\"O}.L. acknowledges useful discussions with Szil\'ard Szalay and Christian Schilling.
}

\appendix
\section{States with maximal entropy}
The eigenvalues of the OP-RDM are denoted by $n_i$ and they satisfy the conditions 
\begin{equation}
0\le n_i\le 1\quad i=1,\ldots,d,
\end{equation}
where $d$ is the number of the modes.
These inequalities determine a convex subset $\cal D$ in ${\mathbb R}^d$.
We define an affine function 
\begin{equation}\label{cons1}
{\cal H}(n_1,\ldots,n_d)=\sum_{i=1}^d n_i-N.
\end{equation}
Due to our normalization of the OP-RDM we have the following constraint
\begin{equation}\label{hom}
{\cal H}(n_1,\ldots,n_d)=0,
\end{equation}
where $N$ is the particle number. The parameters $d$ and $N$ satisfy the inequality $d\geq N\geq 2$.
The state is non-entangled if $d=N$ since the Slater rank is one in this case. In the following we assume $d>N$.

The one-body entanglement entropy is of the following form 
\begin{equation}\label{fun1}
S^{SP}(n_1,\ldots,n_d)=\sum_{i=1}^dh(n_i).
\end{equation}
Since $h(x)$ is concave on the interval $[0,1]$ ($h(0)=0$ and $h(1)=0$) the one-body entanglement entropy $S_{SP}$
is also concave on $\cal D$.
We are looking for the global maximum of $S^{SP}$ on $\cal D$ with the constraint (\ref{hom}). This is a convex optimization problem and so a local maximum is a global one \cite{conv}. The optimal point $(n_1^*,\ldots,n_d^*)$ is determined by the following conditions (see \cite{conv} pp. 141-142) : there exist a $\nu\in{\mathbb R}$  such that 
\begin{equation}
\frac{\partial S^{SP}}{\partial n_i}+\nu=0\quad\quad i=1,\ldots,d
\end{equation}
together with (\ref{hom}). The solutions of these equations are 
\begin{equation}\label{locmax}
n_i^*=\frac{N}{d}\quad i=1,\ldots, d\quad {\rm and}\quad \nu=\log_2\left(\frac{N}{d-N}\right).
\end{equation}
The value of the maximum  is
\begin{equation}\label{ffun}
-N\log_2N+d\log_2d-(d-N)\log_2(d-N).
\end{equation}
In the two-particle case, due to Theorem 1 (Slater decomposition) the maximum value (\ref{ffun}) can not be reached if $d$ is an odd number.  
\section{$M$ dependence of the Slater rank}
In the case of single configuration the Slater rank satisfies (\ref{sr1}) and (\ref{sr2}). We use the proof by case technique for verification. Twice the value of the Slater rank is given by (\ref{n2}) and it depends on  $\bar m$ and $\bar M$. According to the definitions (\ref{meq}) and (\ref{Meq}) 
there are four different inequalities which determine the values of $\bar m$ and $\bar M$. 
These cases are displayed in Table B1. 
\begin{table}
\caption{\label{case1}Possible inequalities for single configurations.}
\begin{indented}
\item[]\begin{tabular}{|c|l|c|}
\br
case&\multicolumn{1}{c|}{inequality}&$\bar M-\bar m$\\
\br
A&$-j_1\leq M-j_2<M+j_2\leq j_1$&$2j_2$\\
B&$-j_1\leq M-j_2\leq j_1\leq M+j_2$&$j_1+j_2-M$\\
C&$M-j_2\leq -j_1\leq M+j_2\leq j_1$&$j_1+j_2+M$\\
D&$M-j_2\leq -j_1<j_1\leq M+j_2$&$2j_1$\\
\br
\end{tabular}
\end{indented}
\end{table}

First we prove (\ref{sr1}). 
Since the symmetry property  $\langle j_1,m_1,j_2,m_2|J,M\rangle=(-1)^{j_1+j_2-J}\langle j_1,-m_1,j_2,-m_2|J,-M\rangle$
the number of exceptional Clebsch-Gordan coefficients in (\ref{j2wf})  $N_{exc}$ are identical for $M$ and $-M$.
This means that according to (\ref{n2}) we have to investigate only the difference $\bar M-\bar m$ to prove (\ref{sr1}).

Let's assume that the $j_1$, $j_2$ and $M$ values are so that the case $A$ happens. If we change the sign of $M$ the quantum numbers 
 $j_1$, $j_2$ and $-M$ remains in the case $A$ so 
the value of $\bar M-\bar m$ is unchanged. If the values of  $j_1$, $j_2$ and $M$ are so that the case 
$B$ occurs then with a sign change of $M$
the new quantum numbers belongs to the case $C$ but according to Table B1 the values of $\bar M-\bar m$  are identical in the considered two cases.
In the other remaining cases, the sign change of $M$ means the following transformations: $C\rightarrow B$ and $D\rightarrow D$.
It is easy to check that the values of $\bar M-\bar m$ are unchanged under these transformations. In this way we proved (\ref{sr1}). 

\begin{table}
\caption{\label{case2}Change of cases if $M$ turn to a larger value $M'$. The dependence 
on $M$ is indicated on the 
quantities $\bar m$ and $\bar M$. The measure of change is $\Delta=(\bar M(M')-\bar m(M'))-(\bar M(M)-\bar m(M))$. }
\begin{indented}
\item[]\begin{tabular}{|c|c|c|c|}
\br
change&$\bar M(M)-\bar m(M)$&$\bar M(M')-\bar m(M')$&$\Delta$\\
\br
$A\rightarrow A$&$2j_2$&$2j_2$&0\\
$A\rightarrow B$&$2j_2$&$j_1+j_2-M'$&$j_1-(M'+j_2)\leq 0$\\
$B\rightarrow B$&$j_1+j_2-M$&$j_1+j_2-M'$&$M-M'<0$\\
$D\rightarrow B$&$2j_1$&$j_1+j_2-M'$&$-j_1-(M'-j_2)\leq 0$\\
$D\rightarrow D$&$2j_1$&$2j_1$&0\\
\br
\end{tabular}
\end{indented}
\end{table}
Next we prove that the Slater rank cannot increase if we increase $M$ ($M\geq 0$). Here we assume that $N_{exc}=0$. Since we consider two different values of $M$ we sign the $M$ dependence of $\bar m$ and $\bar M$. We consider a triplet of quantum numbers  $j_1,j_2$ and $M$ which belongs to a given case of Table B1. If we increase $M$ to $M'$ the new quantum numbers may belong to an another or the same case.
All possible cases are displayed in Table B2. Since $M\ge 0$ we do not have to consider the case $C$. We have to show that 
$\Delta=(\bar M(M')-\bar m(M'))-(\bar M(M)-\bar m(M))\leq 0$. This quantity is shown in the last column of Table B2. We can derive the sign of $\Delta$ using Table B1. Each case $\Delta$ is nonpositive. This means that we proved (\ref{sr2}). 
\end{document}